# Isospin Fractionation in Nuclear Multifragmentation

H.S. Xu[*], M.B. Tsang, T.X. Liu, X.D. Liu, W.G. Lynch, W.P. Tan,

G. Verde, A. VanderMolen, A. Wagner[a], H.F. Xi[b], C.K. Gelbke

*National Superconducting Cyclotron Laboratory and Department of Physics and Astronomy,*

*Michigan State University, East Lansing, MI 48824, USA*

L. Beaulieu, B. Davin, Y. Larochelle[c], T. Lefort, R.T. de Souza, R. Yanez, V. Viola

*Department of Chemistry and IUCF, Indiana University, Bloomington, IN 47405, USA*

R.J. Charity, and L.G. Sobotka

*Department of Chemistry, Washington University, St. Louis, MO 63130, USA*

## Abstract

Isotopic distributions for light particles and intermediate mass fragments have been measured for $^{112}$Sn+$^{112}$Sn, $^{112}$Sn+$^{124}$Sn, $^{124}$Sn+$^{112}$Sn and $^{124}$Sn+$^{124}$Sn collisions at E/A=50 MeV. Isotope, isotone and isobar yield ratios are utilized to obtain an estimate of the isotopic composition of the gas phase, i.e., the relative abundance of free neutrons and protons at breakup. Within the context of equilibrium calculations, these analyses indicate that the gas phase is enriched in neutrons relative to the liquid phase represented by bound nuclei.

[*] On leave from the Institute of Modern Physics, Lanzhou, China
[a] Present address: Institut für Kern- und Hadronenphysik, Forschungszentrum Rossendorf, D-01314 Dresden, Germany
[b] Present address: Convergence Corporation, Houston, Texas
[c] Present address: Laboratoire de Physique Nucléaire, Université Laval, Canada, G1K-7P4.

Nuclear matter is predicted to exhibit a phase transition between a Fermi liquid and a nucleonic gas at temperatures of the order of 5-15 MeV. Analyses of nuclear multifragmentation support the proposition that a region of mixed-phase may be formed in nuclear collisions[1-3]. The role of isospin in such processes has not been extensively investigated; nonetheless, isospin effects may be important [4-6]. For example, recent calculations of two component (neutron and proton) systems predict decreasing critical temperatures for systems of increasing neutron excess, reflecting the fact that a pure neutron liquid probably does not exist [4]. Moreover, the neutron to proton (N/Z) ratio of the gas phase in two component equilibrium calculations is significantly larger than that of the liquid phase [4-7] in contrast to models of fast fragmentation for which the (N/Z) ratio is homogeneous throughout the system [8,9]. Similar equilibrium assumptions lead to predictions of very large (A≈1000) nuclei forming about protons residing within the neutron-rich gas [10] of the inner crust of a neutron star.

Some indications of an isospin fractionation were obtained from the isotopic distributions of light clusters that display a sensitivity to the overall N/Z ratio of the system [11-13]. Comparisons between the N/Z ratios within fragments and the N/Z ratios of nucleons and light clusters at breakup have been complicated, however, by the lack of isotopic resolution of most multifragment detection arrays. New isotopically resolved multifragmentation measurements are presented in this article for systems of different overall N/Z ratio. Using appropriate isotope ratios, the ratio of the free neutron and proton densities at breakup can be extracted with little sensitivity to distortions from secondary decays of particle unstable nuclei. Within the context of the equilibrium assumption, the nucleonic gas is found to be neutron-enriched and the fragments more symmetric at breakup, in qualitative agreement with recent calculations of mixed phase equilibrium within two component lattice gas and mean field approaches [4-6].

To explore these issues, $^{112}$Sn+$^{112}$Sn, $^{112}$Sn+$^{124}$Sn, $^{124}$Sn+$^{112}$Sn and $^{124}$Sn+$^{124}$Sn collisions were studied by bombarding $^{112}$Sn and $^{124}$Sn targets of 5 mg/cm$^2$ areal density with 50 MeV per nucleon $^{112}$Sn and $^{124}$Sn beams from the K1200 cyclotron in the National Superconducting Cyclotron Laboratory at Michigan State University. Isotopically resolved particles with $1 \le Z \le 10$ were detected with nine telescopes of the Large Area Silicon Strip Array (LASSA). Each telescope consists of one 65 μm single-sided silicon strip detector, one 500 μm double-sided silicon strip detector and four 6-cm thick CsI(Tl) scintillators, read out by pin diodes. The 50mm x 50mm lateral dimensions of each LASSA telescope was divided by the strips of the silicon detectors into 256 (≈3x3 mm$^2$) square pixels, providing an angular resolution of about ±0.43°. The center of the LASSA device was located at a polar angle of θ=32° with respect to the beam axis, covering polar angles of 7°≤ θ ≤ 58°. Impact parameter selection was provided by the multiplicity of charged particles [14] measured with LASSA and with 188 plastic scintillator - CsI(Tl) phoswich detectors of the Miniball / Miniwall array [15]; the combined apparatus covered 80% of the total solid angle. Central collisions were selected by gating on the top 4% of the charged-particle multiplicity distribution, corresponding to a reduced impact parameter of $b / b_{max} \le 0.2$ in the sharp cut-off approximation [16]. Previous studies demonstrate that such collisions lead to bulk multifragmentation [17].

Figure 1 shows carbon-isotope particle-identification histograms [18], integrated over all LASSA telescopes for central $^{112}$Sn+$^{112}$Sn (left panel) and $^{124}$Sn+$^{124}$Sn (right panel) collisions. The isotopes from $^{11}$C to $^{15}$C are well resolved. The centroid of the mass distribution for the neutron-rich $^{124}$Sn+$^{124}$Sn system is shifted by about ½ mass unit towards heavier isotopes compared to that for $^{112}$Sn+$^{112}$Sn. The background in the spectrum arises from coincidence summing of signals from light particles and neutrons in the CsI(Tl) scintillators. For particles stopped in the Si detectors, the background is negligible[19].

A basic framework for the extraction of free neutron and proton densities at breakup from isotopic ratios has been provided in ref. [6,20]. Assuming chemical equilibrium [20], one can write the primary multiplicity (before secondary decay of excited states) for an isotope with neutron number, *N,* and proton number, *Z,* in its i-th state as:

$$M_i(N,Z) \propto V(2J_i+1)\cdot e^{(Nm_n+Zm_p+B(N,Z)-E_i^*)/T} \propto V(2J_i+1)\cdot \rho_n^N \cdot \rho_p^Z \cdot e^{B(N,Z)/T}\cdot e^{-E_i^*/T} \quad (1)$$

where *V* is the volume, $\rho_n = M_i(1,0)/V$ and $\rho_p = M_i(0,1)/V$ are the primary free neutron and free proton densities respectively; *B(N,Z),* $J_i$, and $E_i^*$ are the ground state binding energy, spin, and excitation energy of the isotope in the i-th state; $m_n$ and $m_p$ are the neutron and proton chemical potentials and $T$ is the temperature. Secondary decay of excited fragments after breakup introduces corrections to the final yields. The leading order correction arises from particle stable states and is a multiplicative factor, $\tilde{z}_{N,Z}(T) = \sum_{i,stable}(2J_i+1)\cdot e^{-E_i^*/T}$, where the sum is over particle stable states of the fragment. We further assume that the correction due to particle unstable decay can be represented by another multiplicitive factor $f_{N,Z}(T)$:

$$M_{obs}(N,Z) \propto V \cdot r_n^N \cdot r_p^Z \cdot e^{B(N,Z)/T} \cdot \tilde{z}_{N,Z}(T) \cdot f_{N,Z}(T) \qquad (2)$$

In principle, the influence of $\tilde{z}_{N,Z}(T)$ and $f_{N,Z}(T)$ can be assessed by model calculations.

The construction of single and double isotope and isotone ratios provides information about the temperature and the free neutron and proton densities at breakup [20]. From Eq. 2, the multiplicity ratios of two isotopes differing by $k$ neutrons is related to the free neutron density by:

$$\frac{M_{obs}(N+k,Z)}{M_{obs}(N,Z)} = \frac{\tilde{z}_{N+k,Z}(T)\cdot f_{N+k,Z}(T)}{\tilde{z}_{N,Z}(T)\cdot f_{N,Z}(T)} \cdot (r_n)^k \cdot e^{\Delta B/T}, \qquad (3)$$

where *DB=B(N+k,Z)-B(N,Z)* is the binding energy difference between the two isotopes. The multiplicity ratios of two isotones differing by $k$ protons is related to the free proton density by:

$$\frac{M_{obs}(N,Z+k)}{M_{obs}(N,Z)} = \frac{\tilde{z}_{N,Z+k}(T)\cdot f_{N,Z+k}(T)}{\tilde{z}_{N,Z}(T)\cdot f_{N,Z}(T)} \cdot (r_p)^k \cdot e^{\Delta B/T}, \qquad (4)$$

where *DB=B(N,Z+k)-B(N,Z)* is the binding energy difference between the two isotones. The ratio of the multiplicities of neighboring mirror nuclei is related to the ratio of the free neutron and proton densities by:

$$\frac{M_{obs}(Z,N)}{M_{obs}(N,Z)} = \frac{\tilde{z}_{Z,N}(T)\cdot f_{Z,N}(T)}{\tilde{z}_{Z,N}(T)\cdot f_{N,Z}(T)} \cdot \frac{r_n}{r_p} \cdot e^{\Delta B/T} \approx \frac{r_n}{r_p} \cdot e^{\Delta B/T}\, \frac{f_{Z,N}(T)}{f_{N,Z}(T)}, \qquad (5)$$

where $DB=B(Z,N)-B(N,Z)$ is the binding energy difference between the two mirror nuclei and $N=Z-1$. The similarity between the excited particle stable states of the two mirror nuclei leads to an approximate cancellation of the partition functions $\tilde{z}_{N,Z}(T)\,/\,\tilde{z}_{N,Z}(T)\sim 1$.

More relevant for the investigation of isotopic effects is the construction of observables which maximize the sensitivity to isospin effects and minimize the sensitivity to temperature and sequential decays. Extensive studies indicate that distortions from sequential decays are specific to each isotope ratio and depend less on entrance channel [21-23]. To test for an enhanced N / Z ratio of the gas phase, we compare the free neutron and free proton densities obtained from Eq. 3 and 4 for the four Sn+Sn systems. Distortions from sequential decays are minimized by normalizing the isotopic and isotonic ratios in Eqs. 3 and 4 to those of the $^{112}$Sn+$^{112}$Sn system, giving a relative free neutron density,

$$\left(\hat{r}_n\right)^k = \frac{M_{obs}(N_i+k,Z_i)}{M_{obs}(N_i,Z_i)} \Bigg/ \frac{M^{112}_{obs}(N_i+k,Z_i)}{M^{112}_{obs}(N_i,Z_i)} = \left(\frac{r_n}{r^{112}_n}\right)^k, \tag{6}$$

and a relative free proton density,

$$\left(\hat{r}_p\right)^k = \frac{M_{obs}(N_i,Z_i+k)}{M_{obs}(N_i,Z_i)} \Bigg/ \frac{M^{112}_{obs}(N_i,Z_i+k)}{M^{112}_{obs}(N_i,Z_i)} = \left(\frac{r_p}{r^{112}_p}\right)^k. \tag{7}$$

In these expressions, it has been assumed that temperature differences for the four systems (which have nearly the same center of mass energy per nucleon) are negligible[24], enabling the cancellations of both the binding energy factors and the partition functions. Cancellations of the particle unstable feeding corrections are also assumed[21-23].

To test these cancellations, we performed two different Statistical Multifragmentation Model (SMM) calculations [25, 26] with A=248, Z=100 and A=224, Z=100, E* / A=5 MeV. The SMM calculations described in Ref. [25] utilize the available empirical levels and branching ratios for the secondary decay stage, while the calculations in Ref. [26] use parameterizations to calculate both. For simplicity, only $k$=1 isotope and isotone ratios are used to calculate the relative free n and p densities before and after sequential decay. Table I lists the mean relative free nucleon density, $<\hat{r}_n>$ and $<\hat{r}_p>$, before and after secondary decay. The spread of the distribution of ratios is indicated by the standard deviations in the Table. (The standard deviations of the calculations are typically 50% larger than those of the experiment.) Values for $\hat{r}_n$ and $\hat{r}_p$, derived directly from the predicted primary free nucleon multiplicities, are also given. The mean values, $<\hat{r}_n>$ and $<\hat{r}_p>$, obtained from averaging ratios of either the primary or final multiplicities track these calculated relative free nucleon densities at breakup very well. This suggests that experimental values for $<\hat{r}_n>$ and $<\hat{r}_p>$ are robust with respect to distortions from sequential decay and can provide reasonable estimations for the actual relative free neutron and proton densities at breakup.

Experimental values for $<\hat{r}_n>$ and $<\hat{r}_p>$ were extracted for central collisions from isotope yields measured with good statistics and low background and listed in Table II. To reduce the sensitivity to projectile and target remnants, these data were further selected by a rapidity gate of $0.4 \leq y / y_{beam} \leq 0.65$, where y and $y_{beam}$ are the rapidities of the analysed particle and beam, respectively. The resulting mean values, $<\hat{r}_n>$ and $<\hat{r}_p>$, extracted from isotope ratios for the four systems: $^{112}$Sn+$^{112}$Sn, $^{112}$Sn+$^{124}$Sn, $^{124}$Sn+$^{112}$Sn, $^{124}$Sn+$^{124}$Sn are shown in Fig. 2 as a function of the N / Z ratio for the composite system, $(N/Z)_O$. The solid circles and squares denote values for $<\hat{r}_n>$ and $<\hat{r}_p>$ extracted using Eq. 6 and 7 for k=1, the open circles and squares denote corresponding values for k=2 and the stars denote values for k=3. The error bars include both statistical errors, uncertainties arising from background subtraction and systematic errors. The excellent agreement between the extracted values obtained for k=1, 2 and 3 suggests that the factorization into neutron and proton densities in Eq. 1 and the cancellation of the secondary decay corrections in Eq. 6 and 7 are valid. The agreement between the mixed systems, $^{112}$Sn(beam)+$^{124}$Sn(target) and $^{124}$Sn(beam)+$^{112}$Sn(target), with $(N/Z)_O$ =1.36 reflects the fact that the kinematic distortions due to the acceptance of LASSA are negligible.

If the relative concentrations of neutrons and protons were the same throughout each system and the overall nuclear matter density was independent of $(N/Z)_O$, the relative free neutron and proton densities as functions of $(N/Z)_O$ would follow the solid and dashed lines in Fig. 2, respectively. The experimental data shows that as $(N/Z)_O$ increases, the system responds by making the asymmetry of the gas (given by the ratio $<\hat{r}_n>/<\hat{r}_p>$) much greater than the asymmetry of the total system (given by the ratio of solid to dashed lines).

While this analysis of $<\hat{r}_n>$ and $<\hat{r}_p>$ demonstrates the neutron enrichment of the free nucleonic gas, it does not provide absolute values for $r_n/r_p$ [7]. To estimate $r_n/r_p$ using Eq. 5, three pairs of mirror nuclei, $^3$H/$^3$He, $^7$Li/$^7$Be and $^{11}$C/$^{11}$B were analyzed for the $^{112}$Sn+$^{112}$Sn and $^{124}$Sn+$^{124}$Sn systems. The solid points in Figure 3 show the isobaric yield ratios of these mirror nuclei as a function of the binding energy difference, *DB*. The data display a trend similar to Eq. 5, but differ in details, possibly due to Coulomb effects and the secondary decay factor $f_{Z,N}(T)/f_{N,Z}(T)$ which have been neglected. Assuming the correction to the individual ratios due to the secondary decay factor $f_{Z,N}(T)/f_{N,Z}(T)$ to be random and larger than the experimental uncertainties, extrapolating to vanishing *DB* via the solid lines yields values, following Eq. 5, for $r_n/r_p$ of 2.5 for the $^{112}$Sn+$^{112}$Sn system (left panel) and 5.5 for the $^{124}$Sn+$^{124}$Sn system (right panel). Secondary decay calculations using SMM models [25,26], which do not reproduce the detail features of the isotopic distributions, predict that such extrapolations may underestimate $r_n/r_p$ by as much as 30-50%. Alternative extrapolations, denoted by the dashed and dot-dashed lines, provide values for $r_n/r_p$ which range from 1.7 to 3.4 for the $^{112}$Sn+$^{112}$Sn system (left panel) and 2.8 to 8.2 for the $^{124}$Sn+$^{124}$Sn system (right panel). Statistical models with more accurate treatment of the sequential decay may reduce the uncertainties in these estimates. In all cases, the lower limits of these values are significantly larger than $M(\mathrm{n})/M(\mathrm{p})$ =1.24 and 1.48 of the initial system marked by arrows in Fig. 3. This confirms that the gas, consisting of free nucleons, contains proportionately more neutrons than the total system. Both the analyses shown in Figs. 2 and 3 indicate a neutron enhancement for the neutron rich system that is roughly twice that of the neutron deficient system.

A preference for the breakup configurations consisting of neutron rich gas of nucleons and more symmetric fragments is also the outcome expected for statistical multifragment rate equation approaches such as the EES model of ref. [27]. (Note, the fragment production rates in such models are related to the yields of equilibrium multifragment models via detailed balance.) Regardless of whether equilibrium or rate equation approaches are more valid, the observed trends suggest that there is sufficient time in multifragmentation processes for isospin fractionation. The observation is consistent with the conclusions of ref. [28] and the assumptions of refs. [1-6] but it is inconsistent with percolation and cold fragmentation models [8,9]. Isospin effects are also expected for evaporation processes [23,29-31]. However, such studies should be undertaken within the context of compound nuclear decay theory [31,32] and are beyond the scope of this short letter.

In summary, we have measured the isotope distributions from Z=1 to Z=8 particles emitted in four different Sn+Sn reactions ranging from initial isospin of 1.24 to 1.48. The nucleon density extracted from the gas phase using isotope, isotone and isobar ratios suggest that the gas phase is more enriched in neutrons than the liquid phase represented by bound nuclei, consistent with the predicted partial fractionation of nucleon components in the liquid gas phase transition. The neutron enrichment is much more enhanced in the collisions of neutron-rich systems as compared to the collisions of neutron-deficient system.

The authors wish to acknowledge many fruitful discussions with Drs. P. Danielewicz, W.A. Friedman and S. Pratt. We are especially grateful to Dr. Jan Toke for his help in the data acquisition. This work is supported by the National Science Foundation under Grant No. PHY-95-28844.

## FIGURE CAPTIONS:

Figure 1: Carbon isotope particle identification histograms obtained for $^{112}$Sn+$^{112}$Sn (left panel) and $^{124}$Sn +$^{124}$Sn (right panel) reactions.

Figure 2: The mean relative free neutron and free proton density (Eq. 6 and 7) as a function of $(N/Z)_O$. The solid and dotdashed lines are the expected n-enrichment and p-depletion with the increase of isospin of the initial systems.

Figure 3: Isobar ratios for three mirror nuclei obtained from the $^{112}$Sn+$^{112}$Sn (left panel) and $^{124}$Sn+$^{124}$Sn (right pane) reactions. See text for explanation of the lines.

**Table I:** Relative free proton and neutron densities (Eq. 6 and 7) from two model calculations before and after sequential decays and from primary free proton and neutron multiplicites.

| $\hat{r}_n = \dfrac{M^{124}(\text{n})/124}{M^{112}(\text{n})/112}$ | $<\hat{r}_n>$ (primary) | $<\hat{r}_n>$ (final) | $\hat{r}_n = \dfrac{M^{124}(\text{p})/124}{M^{112}(\text{p})/112}$ | $<\hat{r}_p>$ (primary) | $<\hat{r}_p>$ (final) | Models [Ref.] |
|---|---|---|---|---|---|---|
| 1.70 | 1.65±0.10 | 1.60±0.13 | 0.50 | 0.49±0.03 | 0.60±0.15 | SMM1 [25] |
| 1.54 | 1.73±0.24 | 1.71±0.17 | 0.44 | 0.54±0.09 | 0.61±0.09 | SMM2 [26] |

**Table II**: List of isotope and isotone ratios used in the present work.

| **Isotope Ratios** ΔA=1 | ΔA=2 | ΔA=3 | **Isotone Ratios** ΔZ=1 | ΔZ=2 | ΔZ=3 |
|---|---|---|---|---|---|
| d / p | | | $^{3}$He / d | | |
| t / d | t / p | | $^{4}$He / t | | |
| $^{4}$He / $^{3}$He | $^{6}$He / $^{4}$He | $^{6}$He / $^{3}$He | $^{7}$Li / $^{6}$He | | |
| $^{7}$Li / $^{6}$Li | | | $^{7}$Be / $^{6}$Li | | |
| $^{8}$Li / $^{7}$Li | $^{8}$Li / $^{6}$Li | | $^{9}$Be / $^{8}$Li | | |
| $^{9}$Li / $^{8}$Li | $^{9}$Li / $^{7}$Li | $^{9}$Li / $^{6}$Li | $^{10}$B / $^{9}$Be | $^{10}$B / $^{8}$Li | |
| $^{10}$Be / $^{9}$Be | $^{9}$Be / $^{7}$Be | $^{10}$Be / $^{7}$Be | $^{11}$B / $^{10}$Be | $^{11}$B / $^{9}$Li | |
| $^{11}$B / $^{10}$B | | | $^{12}$C / $^{11}$B | $^{12}$C / $^{10}$Be | $^{12}$C / $^{9}$Li |
| $^{12}$B / $^{11}$B | $^{12}$B / $^{10}$B | | $^{13}$C / $^{12}$B | | |
| $^{13}$B / $^{12}$B | $^{13}$B / $^{11}$B | $^{13}$B / $^{10}$B | $^{14}$C / $^{13}$B | | |
| $^{13}$C / $^{12}$C | | | $^{14}$N / $^{13}$C | $^{14}$N / $^{12}$B | |
| $^{14}$C / $^{13}$C | $^{14}$C / $^{12}$C | | $^{15}$N / $^{14}$C | $^{15}$N / $^{13}$B | |
| $^{15}$N / $^{14}$N | | | $^{16}$O / $^{15}$N | $^{16}$O / $^{14}$C | $^{16}$O / $^{13}$B |
| $^{16}$N / $^{15}$N | $^{16}$N / $^{14}$N | | $^{17}$O / $^{16}$N | $^{17}$O / $^{15}$C | |
| $^{17}$N / $^{16}$N | $^{17}$N / $^{15}$N | $^{17}$N / $^{14}$N | $^{18}$O / $^{17}$N | | |
| $^{17}$O / $^{16}$O | | | | | |
| $^{18}$O / $^{17}$O | $^{18}$O / $^{16}$O | | | | |

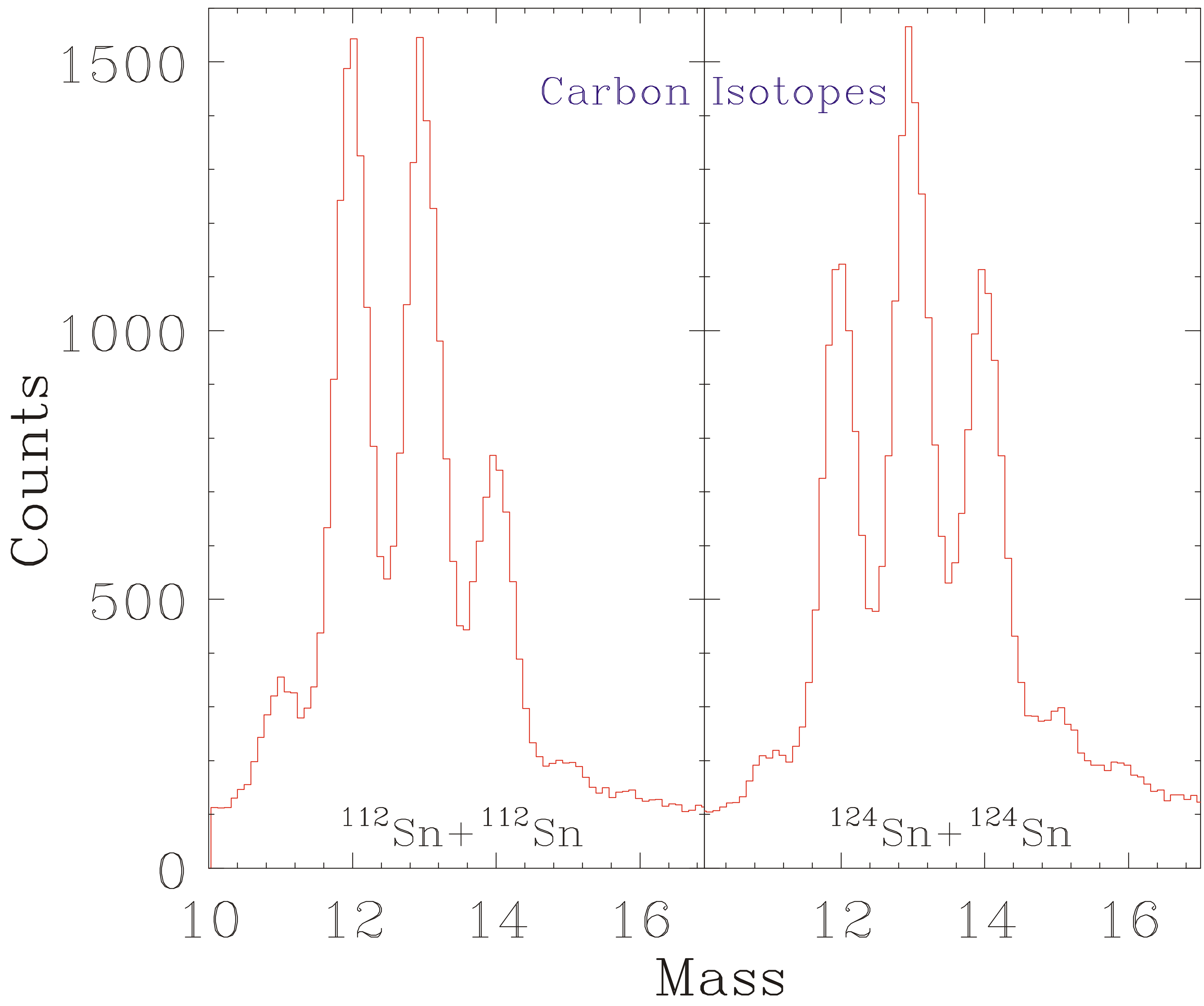

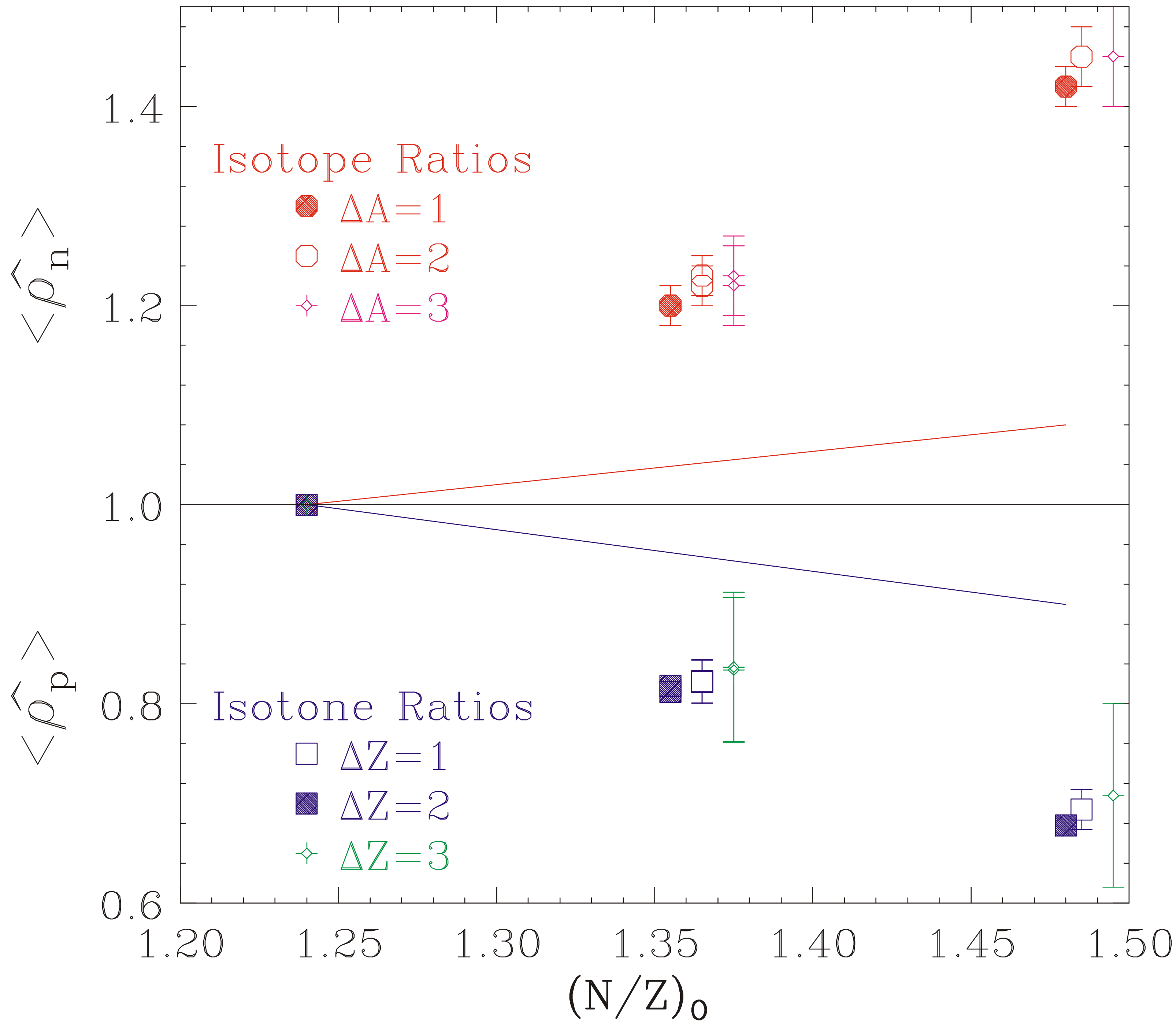

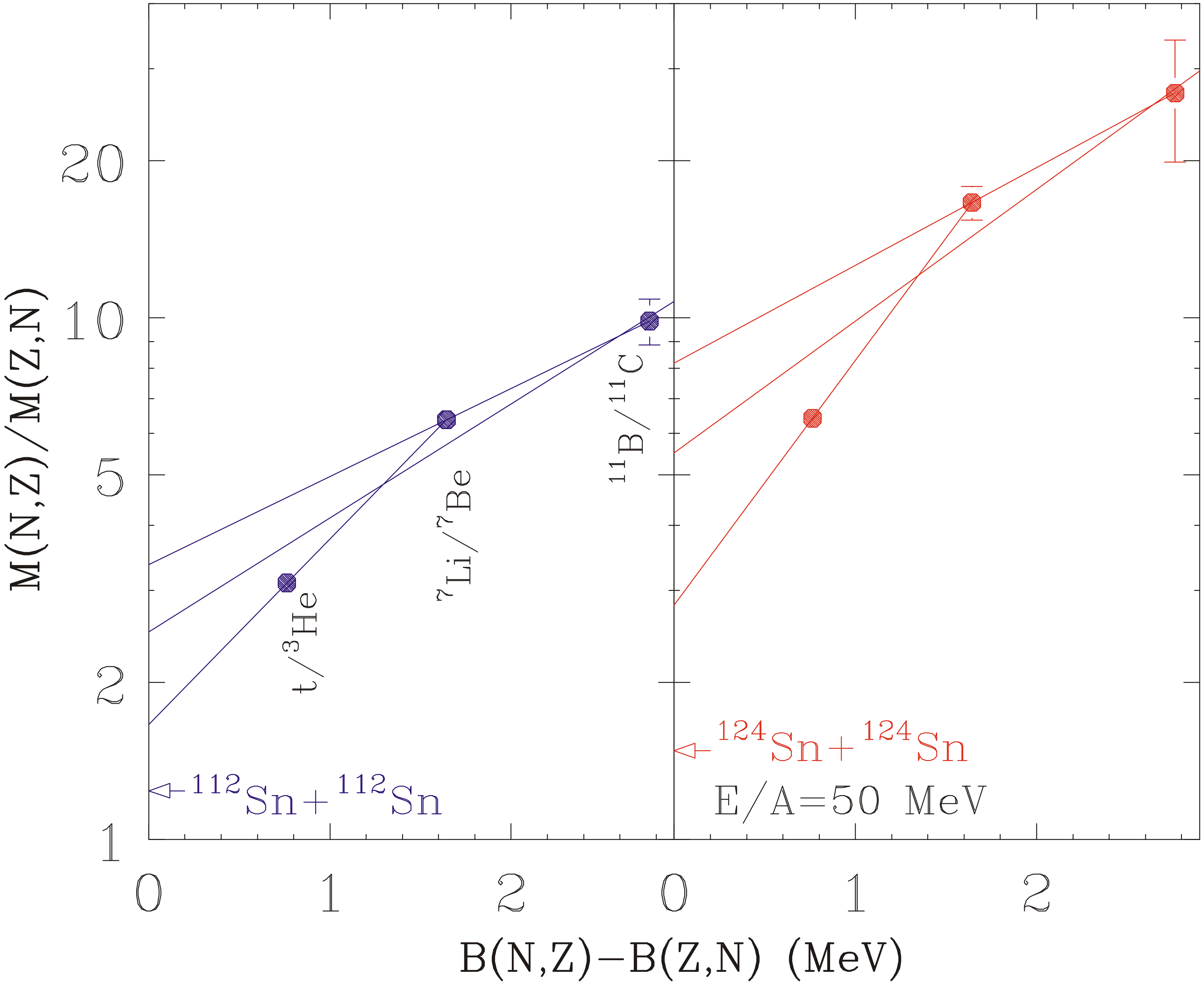